\pdfoutput=1

\documentclass[aps,prl,10pt,twocolumn,superscriptaddress]{revtex4-1}

\usepackage{times, mathrsfs, amsmath, amsfonts, cancel, color, theorem, bbm, amssymb, latexsym}

\newcommand{\Tr}{\mbox{Tr}}

\newcommand{\bra}[1]{\langle #1 |}
\newcommand{\ket}[1]{| #1 \rangle}
\newcommand{\ketbra}[1]{| #1 \rangle\langle #1 |}
\newcommand{\braket}[1]{\langle #1 \rangle}

\newtheorem{definition}{Definition}

\newtheorem{theorem}[definition]{Theorem}

\newtheorem{conjecture}[definition]{Conjecture}

\begin{document}

\date{\today}
\title{Masking quantum information is impossible}

\author{Kavan Modi}
\email{kavan.modi@monash.edu}
\affiliation{School of Physics \& Astronomy, Monash University, Victoria 3800, Australia}

\author{Arun Kumar Pati}
\email{akpati@hri.res.in}

\author{Aditi Sen(De)}
\email{aditi@hri.res.in}

\author{Ujjwal Sen}
\email{ujjwal@hri.res.in}

\affiliation{Quantum Information and Computation Group,\\
Harish-Chandra Research Institute, HBNI, Chhatnag Road, Jhunsi, 
Allahabad 211 019, India}


\begin{abstract}
Classical information encoded in composite quantum states can be completely hidden from the reduced subsystems and may be found only in the correlations. Can the same be true for quantum information? If quantum information is hidden from subsystems and spread over quantum correlation, we call it as masking of quantum information. We show that while this may still be true for some restricted sets of non-orthogonal quantum states, it is not possible for arbitrary quantum states. This result suggests that quantum qubit commitment -- a stronger version of the quantum bit commitment is not possible in general. Our findings may have potential applications in secret sharing and future quantum communication protocols.
\end{abstract}

\maketitle

In a quantum world, information encoded in arbitrary pure quantum states cannot be copied  perfectly, a result known as the no-cloning theorem~\cite{zurek, dieks, yuen, cloningexp}. It plays an important role in several quantum information processing tasks like  quantum key distribution~\cite{qkdRMP} and quantum teleportation~\cite{teleportation}. It was also shown that impossibility of copying pure states can be extended to arbitrary density matrices resulting in the no-broadcasting theorem~\cite{nobroadcast, hen}. On the other hand, deleting quantum information in a closed system is also known to be impossible~\cite{pati}. All these no-go theorems  are consequences of the linearity and the unitarity of quantum mechanics. If we are given a set of non-orthogonal states, unitarity prohibits cloning or deleting of quantum states. A stronger version of the no-cloning theorem states that quantum copying machine exists only when the blank state already possess the full information of the input state~\cite{jozsa}. Together with the no-deleting theorem, it gives a permanence to quantum information--a notion that is only true for quantum information which does not hold in a classical world (for other no-go theorems, see~\cite{akp, nosplit, nopar, indra} and in particular the no-go theorems on quantum bit commitment~\cite{nobitcom, nobit}). 

Not surprisingly, the no-cloning and the no-deleting theorems are closely connected to the conservation of information and the second law of thermodynamics~\cite{sen, horo}. This gives us an impression that quantum information is truly robust in some sense. However, we also know that when a quantum system interacts with the external world, it may loose its coherence and  information from a quantum state  may disappear completely from the original system in some extreme cases. Can such phenomena indicate loss of information like Maxwell's demon~\cite{maxwelldemon}? However, using the linearity and the unitarity of quantum mechanics, one can prove that whenever there is loss of information from one system, there must be appearance of the same in some subspace of the environment~\cite{sam}. This is known as the no-hiding theorem. It shows that there is no information loss in reality and conservation of quantum information in its full generality holds. A recent experiment by using nuclear magnetic resonance shows that indeed information is conserved when a qubit undergoes state randomisation and can be fully recovered from the ancillary states by applying local unitary operator in the ancillary Hilbert space~\cite{anil}.

Let us now consider an example of hiding classical information by using quantum correlation of a two-party state. Suppose, we encode a single bit of classical information in two orthogonal entangled states where the encoding map is given by $\ket{0} \rightarrow \tfrac{1}{\sqrt{2}}(\ket{00} + \ket{11})$ and $\ket{1} \rightarrow \tfrac{1}{\sqrt{2}} (\ket{00} - \ket{11})$. If we look at states of both the subsystems, it has no information about the classical bit. Here, we can say that although classical information is actually hidden from both the subsystems, it is spread over quantum correlation of the encoded states. 

In this paper, we deal with the  encoding of quantum information in an arbitrary \emph{composite} quantum state. We ask the question: \emph{can quantum information be hidden from both the subsystems and remain only in the correlation?} If so, then somehow quantum information gets spread over the `spooky' correlation and remains invisible to both the subsystems that are possessed by the local observers.  We call this spreading of quantum information over quantum correlations as `masking' quantum information~\cite{eita-spooky, eita-o-spooky}. We prove that such masking is not possible for arbitrary quantum states, although we have already seen that it is possible for classical information to be masked. For some restricted classes of quantum states, however, masking is possible. Indeed we show that there are sets of quantum states whose information we can mask, which  are continuous and contain non-orthogonal states.

Our result has immediate applications in quantum bit commitment~\cite{nobitcom} and quantum secret sharing protocols~\cite{secretsharing, buzek, besh, imoto, lo}. In quantum bit commitment, the receiver (Bob) is blind to the sender's (Alice's) committed bit, and this is translated to the condition that the subsystem of the encoded entangled state has no information about the committed bit. We propose a quantum {\it qubit} commitment where Alice is committed to a qubit chosen from an alphabet of qubit states, and later she wants to convince Bob that she had indeed chosen one of the states from that set. From our result, it follows that such a scheme is not possible, in general. Since the classical bit is a special case of a qubit (obtained by passing the qubit through a dephasing channel), no bit commitment also follows from our theorem. Moreover, our results imply that the set of states which can be masked are useful for quantum secret sharing and may have applications in future quantum communication protocols.

\emph{Masking quantum information.---} We begin by formally defining masking of quantum information.

\begin{definition} 
An operation $\mathcal{S}$ is said to mask quantum information contained in states $\{ \ket{a_k}_A \in \mathcal{H}_A \}$ by mapping them 
to states $\{\ket{\Psi_k}_{AB} \in \mathcal{H}_{A} \otimes \mathcal{H}_{B}\}$ such that all the marginal states of $\ket{\Psi_k}_{AB}$ are identical, i.e., 
\begin{gather}\label{traeq}
\rho_A = \Tr_B( \ket{\Psi_k}_{AB} \bra{\Psi_k}) \quad \mbox{and} \quad
\rho_B = \Tr_A( \ket{\Psi_k}_{AB} \bra{\Psi_k})
\end{gather}
have no information about the value of $k$.
\end{definition}

We call such a machine $\mathcal{S}$ as the masker. Since the action of the masker is a physical process, it can be modelled by a unitary operator 
$U_\mathcal{S}$ on the system $A$ plus an ancillary system $B$, given by
\begin{gather}
\mathcal{S}: U_{\mathcal{S}}\ket{a_k}_A \otimes \ket{b}_B =
\ket{\Psi_k}_{AB}.
\end{gather}
This is a linear transformation and it preserves orthogonality. Moreover, if $\mathcal{S}$ can mask information in a set of states $\{\ket{a_k}\}$, then it can mask the information contained in a state whose density matrix can be expressed as a linear combination of density matrices $\{\ketbra{a_k}\}$. Furthermore, it is important to require that neither $A$ nor $B$ contain the information of the initial state. Otherwise a simple application of \textsc{swap} gate will mask the information for $A$ by simply transferring it to $B$. Therefore, we demand that masked information solely lies in the correlations between $A$ and $B$. This means that the final state must be an entangled pure state and the marginal states $A$ and $B$ contain exactly the same information.

We now prove that it is impossible to mask the information in any arbitrary quantum state.  This theorem is in the same spirit as the no-cloning  and no-deleting theorems~\cite{zurek, dieks, pati}. However, we will show below that the set of maskable states is much richer than the set of states which can be cloned and deleted.

\begin{theorem}
No masker can mask all states of a qubit in ${\cal H}^2$.
\end{theorem}

\noindent\emph{Proof.} Let us assume that $\mathcal{S}$ can mask all states of a qubit in $\mathcal{H}_A$. Let $\{\ket{k}\}_{k=0}^1$ be an orthonormal basis on $\mathcal{H}_A$ and the action of the masker gives us $\mathcal{S}: \ket{k}\to \ket{\Psi_k}$, where $\ket{\Psi_k}$ are also orthonormal. Now, let us express an arbitrary quantum state in terms of the basis elements of an orthonormal basis as $\ket{a}=\sum_{k=0}^{1} \alpha_k \ket{k}$. We now assume that   $\ket{a}$ can be masked, i.e.,
\begin{gather}
\ket{a}= \alpha_1 \ket{0} +  \alpha_2 \ket{1} \, \to \,
\ket{\Psi}= \alpha_1 \ket{\Psi_0} + \alpha_2 \ket{\Psi_1},
\end{gather}
where  $|\alpha_1|^2  + |\alpha_2|^2 =1$. Next, we take partial trace with respect to either $A$ or $B$ to get 
\begin{align}
\Tr_X [\ketbra{\Psi}] =  \rho_Y  +& \Tr_X (\alpha_1 \alpha_2^* \ket{\Psi_0} \bra{\Psi_1}) 
\nonumber \\
&+ \alpha^*_1 \alpha_2  \Tr_X (\ket{\Psi_1} \bra{\Psi_0}),
\end{align}
where $\{X,\, Y\} \in \{A,\, B\}$ and $X \ne Y$. The last equation satisfy the masking conditions if the off-diagonal terms vanish, namely
\begin{gather}
\alpha_1 \alpha_2^* \Tr_X (\ket{\Psi_0} \bra{\Psi_	}) + \alpha^*_1 \alpha_2  \Tr_X (\ket{\Psi_1} \bra{\Psi_0})= 0,
\end{gather}
for arbitrary $\alpha_1$ and $\alpha_2$. It implies that we have
\begin{gather}\label{traeq01}
\Tr_X(\ket{\Psi_0} \bra{\Psi_1}) = \Tr_X (\ket{\Psi_1} \bra{\Psi_0} )=0.
\end{gather}

We will now show that the above conditions cannot be satisfied for an arbitrary qubit. To prove this, we will use a result, given  in Ref.~\cite{hardy}, for writing  two orthogonal states, which are given by
\begin{align}\label{eq:walgate}
&\ket{\Psi_0} = \ket{\mu } \otimes \ket{ 0} + \ket{\nu} \otimes  \ket{ 1} \quad \mbox{and} \notag \\
&\ket{\Psi_1} = \ket{\mu_\perp} \otimes \ket{ 0} + \ket{\nu_\perp} \otimes \ket{ 1},
\end{align}
where $\ket{\mu}$ and $\ket{\nu}$  are not  necessarily normalized and not mutually orthogonal while $\ket{\mu}$ ($\ket{\nu}$) and $\ket{ \mu_\perp}$  ($\ket{ \nu_\perp}$) are mutually orthogonal. Since the masked states are orthogonal, we will use  this decomposition. Let us now compute the partial traces with respect to $B$ explicitly. We have
\begin{align}
&\Tr_B[\ketbra{\Psi_0}]= \ketbra{\mu} +\ketbra{\nu}, \label{tra0}\\
&\Tr_B[\ketbra{\Psi_1}]= \ketbra{\mu_\perp} + \ketbra{\nu_\perp}, \label{tra1}\\
&\Tr_B[\ket{\Psi_0}\bra{\Psi_1}]=\ket{\mu}\bra{\mu_\perp}+\ket{\nu}\bra{\nu_\perp}. \label{tra01}
\end{align}
Using Eq.~\eqref{traeq}, we get 
$$\ketbra{\mu} + \ketbra{\nu}= \ketbra{\mu_\perp} + \ketbra{\nu_\perp}.$$ 
The expectation value of the above equation with respect to $\ket{\mu}$ gives
\begin{gather}\label{controdiction}
|\braket{\mu|\mu}|^2+|\braket{\nu|\mu}|^2=|\braket{\nu_\perp | \mu}|^2.
\end{gather}
Now using Eq.~\eqref{traeq01} and taking the expectation value of the operator in Eq.~\eqref{tra01} with respect to $\ket{\mu}$, we get
\begin{gather}
\braket{\mu|\nu}\braket{\nu_\perp|\mu}=0,
\end{gather}
which implies either $\braket{\mu|\nu}=0$ or $\braket{\nu_\perp|\mu}=0$.  But in either case that makes Eq.~\eqref{controdiction} into
\begin{gather}
|\braket{\nu_\perp|\mu}|^2= |\braket{\mu|\mu}|^2 \quad \mbox{or} \quad |\braket{\nu|\mu}|^2=-|\braket{\mu|\mu}|^2.
\end{gather}
The latter is a contradiction, while in the former case, we have $\ket{\nu_\perp} = {\rm e}^{i \phi} \ket{\mu}$. Using this fact and taking the inner product in Eq.~\eqref{tra01} with $\bra{\mu}$ and $\ket{\mu_\perp}$, we obtain
\begin{gather}
\braket{\mu|\Tr_B(\ket{\Psi_0} \bra{\Psi_1})|\mu_\perp} = \braket{\mu|\mu} \braket{\mu_\perp|\mu_\perp} = 0.
\end{gather}
Last equation means that either $\ket{\mu}=0$ or $\ket{\mu_\perp}=0$. If so, in either case, the states in Eq.~\eqref{eq:walgate}  are not entangled, implying that  the states of $A$ and $B$ can be simply swapped. This is a contradiction. Therefore, arbitrary qubits can not be masked.
\hfill $\blacksquare$

Above we have shown that arbitrary two-dimensional quantum states can not be masked. We will now show that this Theorem holds in arbitrary dimensions. Interestingly, note that the proof that is given below in arbitrary dimension is different  than that in Theorem 2. In particular, the Theorem 3 below uses the Schmidt decomposition, instead of the decomposition of two orthogonal states~\cite{hardy}. 

\begin{theorem}
An arbitrary quantum state cannot be masked. 
\end{theorem}

\noindent{\emph Proof.}
Let us assume that a machine can mask two states, $\ket{s_0}$ and $\ket{s_1}$. Let $\ket{s_0} \rightarrow \ket{\Psi_0}$ and $\ket{s_1} \rightarrow \ket{\Psi_1}$, where \(\ket{\Psi_0}\) and \(\ket{\Psi_1}\) are shared by two parties, \(A\) and \(B\)  in ${\cal H}_A^{d_A} \otimes {\cal H}_B^{d_B}$. Then the superposition states,  $\{\mu \ket{s_0} + \nu \ket{s_1}\}$ with arbitrary coefficients satisfying $|\mu|^2+|\nu|^2 =1$, can also be masked by the same machine.

Since $\ket{\Psi_0}$ and $\ket{\Psi_1}$ are purifications of 
$\rho_A^{(0)}$ and $\rho_A^{(1)}$ respectively, and $\rho_A^{(0)} = \rho_A^{(1)}$, we can write them in Schmidt decomposition as
\begin{gather}
\label{eq:schmidt}
\ket{\Psi_0}= \sum_k \sqrt{\lambda_k} \ket{a_k}  \ket{ b_k^{(0)}}, \quad \ket{\Psi_1}= \sum_k \sqrt{\lambda_k} 
\ket{a_k} \ket{ b_k^{(1)}},
\end{gather}
where $\lambda_k$ are the eigenvalues of the reduced density matrices, which has eigenvectors $\{\ket{a_k}\}_{k=1}^{d}$, with $d=\min\{d_A,d_B\}$. Note that the eigenvectors are orthonormal, i.e., $\braket{a_k | a_l} = \delta_{kl}$. Similarly $\{\ket{b_k^{(0)}}\}$ is also a set of orthonormal vectors, as is $\{\ket{b_k^{(1)}}\}$. 

The masking condition means that the reduced states of $B$ must be the same, i.e., 
\begin{gather}
\rho_B = \Tr_A [\ket{\Psi_0}\bra{\Psi_0} ] = \Tr_A [\ket{\Psi_1}\bra{\Psi_1} ].
\end{gather}
Let us now assume that we can mask the superposition state.  It then implies that we can mask $\mu \ket{\Psi_0} + \nu \ket{\Psi_1}$, due to the linearity of the masker. Taking the partial trace with respect to $A$, we have 
\begin{align}
\label{eq:condition1}
\rho_B =& |\mu|^2 \Tr_A [\ket{\Psi_0}\bra{\Psi_0} ]
+ |\nu|^2 \Tr_A [\ket{\Psi_1}\bra{\Psi_1} ] \notag\\
&+\mu \nu^* \Tr_A [\ket{\Psi_0}\bra{\Psi_1} ]
+\mu^* \nu \Tr_A [\ket{\Psi_1}\bra{\Psi_0} ] \nonumber \\
=& \rho_B +\mu \nu^* \Tr_A [\ket{\Psi_0}\bra{\Psi_1} ]
+\mu^* \nu \Tr_A [\ket{\Psi_1}\bra{\Psi_0} ].
\end{align}
The masking condition demands that the cross terms in Eq.~\eqref{eq:condition1} must vanish and we get
\begin{gather}\label{eq:cond}
\mu \nu^* \Tr_A [\ket{\Psi_0}\bra{\Psi_1} ]
+\mu^* \nu \Tr_A [\ket{\Psi_1}\bra{\Psi_0} ] = 0.
\end{gather}
Using Eq.~\eqref{eq:schmidt}, Eq.~\eqref{eq:cond} reduces to
\begin{gather}
\label{eq:operator1}
\mu \nu^* \sum_k \lambda_k \ket{b_k^{(0)}}\bra{b_k^{(1)}}
+\mu^* \nu \sum_k \lambda_k
\ket{b_k^{(1)}} \bra{b_k^{(0)}}= 0.
\end{gather}
There are no cross terms like $\ket{b_j^{(0)}}\bra{b_k^{(1)}}$ because of orthonormality of vectors $\{\ket{a_k}\}$. 
By taking the expectation value of Eq.~\eqref{eq:operator1} with $\ket{b_j^{(0)}}$, we get 
\begin{gather}
\lambda_j \left(\mu \nu^*   \braket{b_j^{(1)}|b_j^{(0)}} +
\mu^* \nu  \braket{b_j^{(0)}|b_j^{(1)}} \right)= 0.
\end{gather}
Since we can always choose $\lambda_j > 0$,  the solutions are either $\mu=0$, or $\nu=0$, or $\braket{b_j^{(1)} |b_j^{(0)}} =0$, or $\mu \nu^* \braket{b_j^{(1)} |b_j^{(0)}}$ is purely imaginary for all choices of $j$, implying restrictions on the choices of  $\mu, \, \nu$. Therefore, an arbitrary qudit state cannot be masked.  
\hfill $\blacksquare$

The no-local broadcasting theorem~\cite{piani} cleanly differentiates between classical information, which can be copied, and quantum information, which cannot be copied. Such is not the case with masking of quantum information because there are a continuous family of quantum states that can be masked. This finding blurs the boundary that separates the quantum and classical worlds. We now define such a masker $\mathcal{S}^\sharp$ and identify the set of states  that  $\mathcal{S}^\sharp$ can mask. Let $\{\ket{k}\}_{k=1}^d$ be an orthonormal basis in $\mathcal{H}_A$. The joint unitary operation corresponding to the masker $\mathcal{S}^\sharp$ is given by
\begin{gather}\label{eq:spk}
\mathcal{S}^\sharp: \ket{k\, b}_{AB} \to \ket{k\, k}_{AB}.
\end{gather}

\begin{theorem}\label{prop:spook}
Masker $\mathcal{S}^\sharp$ can mask the quantum information if it acts on a state belonging to a family of states on the great hyper-disk whose extremal states are $\{\ket{a} = \tfrac{1}{\sqrt{d}} \sum_k {\rm e}^{i \phi_k} \ket{k}\}$, with the quantum information encoded in the continuous parameters $\{\phi_k \in [-\pi,\pi]\}$.
\end{theorem}

\noindent\emph{Proof.} Using $\mathcal{S}^\sharp$ in Eq.~\eqref{eq:spk}
we have
\begin{gather}
\mathcal{S}^\sharp \ket{a \, b} 
=\frac{1}{\sqrt{d}} \sum_k {\rm e}^{i \phi_k} \ket{k\, k} 
= \ket{\Psi}.
\end{gather}
Partial trace with either system yields a maximally mixed state. By convexity we can mask all states on the great hyper-disk. 
\hfill $\blacksquare$

The masker $\mathcal{S}^\sharp$ can also mask any family of states $\{\ket{\tilde{a}} = \sum_k {\rm e}^{i \phi_k} r_k \ket{k}\}$ that have the amplitudes $r_k$ in common. In fact, above we have only considered the special case where $r_k = 1/\sqrt{d} \; \forall \; k$. Theorem~\ref{prop:spook} can be proven in this more general case with minor modifications. The key difference is that the marginal states for this case are diagonal in the basis $\ket{k}$ with eigenvalues $|r_k|^2$. Therefore the marginals do not contain any information about the phase. It may be noted here that the set of states on the great hyper-disk is of zero measure in the set of all states.

In the scenario that we have considered until now, the encoding states are pure states. We can consider the question whether a similar analysis is possible in the situation where the masker takes pure states to mixed states. This is an open dynamic, and to ensure that the masking is complete, we must require that the local parts of the environment states do not carry any information about the input states. We now further require that the environment states and the system states have vanishing quantum correlations~\cite{Akbar}. This is indeed possible. In particular, we can replace the encoding states in the proof of Theorem 4 by \(\frac{1} {\sqrt{d}} \sum_k e^{i \phi_k} \ket{kkkk}\), where the first two parties represent one party, say Alice, and her environment (call them \(A\) and \(E_A\)), while the last two parties represent the other party, say Bob, and his environment (call them \(B\) and \(E_B\)). In this case, reduced density matrices of the system  as well as the environment  are classically correlated, having zero quantum correlations, and  the masking works as before. Note, however, that the state in the \(AE_A:BE_B\) partition is still entangled.

\begin{conjecture} Based on the structure of the masker $\mathcal{S}^\sharp$ in Eq.~\eqref{eq:spk}, we conjecture that the maskable states corresponding to any masker belong to some disk.
\end{conjecture}


{\em No qubit commitment.--} In a bit commitment protocol, Alice commits to a bit $0$ or $1$ and later she provides Bob, classical or quantum information, that reveals the committed bit. An ideal bit commitment protocol should ensure Bob that Alice is indeed committed to her initial bit and Bob can learn no information about the committed bit before the opening phase. However, the entanglement based cheating strategy makes any quantum bit commitment protocol impossible in the nonrelativistic domain (cf.~see~\cite{Adlam2015} and references therein). To recall, suppose that Alice prepares two two-particle quantum states $\ket{\Psi_0}$ and $\ket{\Psi_1}$ corresponding to bit $0$ or $1$,  keeps one particle, and sends the other to Bob. As Bob has no information about $0$ or $1$, this makes the reduced density matrix $\rho_B= \Tr_A{\ket{\Psi_0}\bra{\Psi_0}}=  \Tr_A{\ket{\Psi_1}\bra{\Psi_1}}$. This condition then implies that $\ket{\Psi_0} = \sum_i \sqrt{\lambda_i} \ket{a_i^{0}} \ket{b_i}$ and $\ket{\Psi_1} = \sum_i \sqrt{\lambda_i} \ket{a_i^{1}} \ket{b_i}$. However, $\ket{\Psi_0} = U_A \otimes I_B \ket{\Psi_1}$ as they differ only by a local change of basis. This is the key to cheating, because during the unveiling stage, Alice can decide to do nothing or apply a local unitary on her particle. Thus, she can always cheat on her committed bit. 

Our results can have application in a {\it no-qubit commitment} protocol where Alice commits to a qubit from certain set (that can potentially also contain nonorthogonal states), instead of a bit, and later unveils to Bob that she has indeed committed to that qubit. Suppose, Alice wants to commit to an arbitrary state of a qubit from a set $\{ \ket{\psi} \}$. Then she needs to prepare an entangled state $\ket{\Psi(\psi)}$ for each $\ket{\psi}$ with the condition that $\rho_B= \Tr_A{\ket{\Psi}\bra{\Psi}}$ is independent of  $\ket{\psi}$. But, by the no-masking theorem, it is impossible to achieve this if the set $\{ \ket{\psi} \}$ is the set of all states. Hence, committing to an arbitrary qubit or qudit is impossible. However, there  is a trivial way to commit, i.e., Alice encodes $\ket{\psi}$ in a product state $\ket{\psi} \ket{0}$ and $\rho_B$ has no 
information about $\ket{\psi}$. But in this encoding, it is trivial to cheat. In the second scenario, we ask if it is possible to commit to two quantum states and have a qubit commitment protocol. By our result, it is possible to mask two quantum states and hence Alice can ensure that committed qubit or qudit is blind to Bob. But again by entanglement cheating strategy, Alice can always cheat.  The usual no bit commitment proof may be considered as a dephased version of no qubit commitment protocol. 

To illustrate the cheating strategy in the qubit commitment protocol, imagine that Alice commits a qubit state chosen from two non-orthogonal states $\ket{\psi_1}$ and $\ket{\psi_2}$, where
\begin{gather}
\ket{\psi_1} = \frac{1}{2} (\ket{0} + \ket{1}), \quad
\ket{\psi_2} = \frac{1}{2} (\ket{0}  + e^{i\phi} \ket{1})
\end{gather}
Note that these two states can be masked  by a map given by
\begin{gather}
\ket{\psi_1} \rightarrow \frac{1}{2} (\ket{00} + \ket{11}), \quad
\ket{\psi_2} \rightarrow \frac{1}{2} (\ket{00} +
e^{i\phi} \ket{11})
\end{gather}
She keep one of the qubit and sends the other qubit to Bob.  The fact that these two states have the same local reduced state, Bob does not know which qubit Alice has actually committed to. Alice's task is to convince Bob that she has indeed committed to one of these two non-orthogonal states. However, this is not possible. Even if she has committed to a qubit chosen from $\{\ket{\psi_1}, \ket{\psi_2} \}$ at the unveiling phase, Alice can apply a local unitary transformation that can change  $\ket{\psi_1} \leftrightarrow  \ket{\psi_2} $. This can be achieved by a unitary operator that maps $\ket{0} \leftrightarrow  \ket{0}$ and $\ket{1} \leftrightarrow  e^{i\phi}  \ket{1}$. Therefore, even if Alice can choose a qubit state from a set that can be masked, it is possible to cheat at the opening stage of the protocol.

It should be stressed that it is not possible to derive the no qubit commitment result from the no bit commitment one. This is because even though there is more information to be hidden by Alice, there is also more information to be extracted by Bob, and there is more space in the Hilbert space for hiding, as we are considering non-orthogonal states for encoding, unlike orthogonal states for bit commitment. Moreover, 
we are hiding quantum information instead of classical information.
The comparison is similar to that in quantum error correction or in fault tolerant quantum computation versus their classical sisters. Focusing on error correction, we know that classical error correction exists even though classical error tries to frustrate/destroy classical information. Quantum noise is far richer and destroys quantum information through far richer channels. However, there are also far richer ways of correcting errors in the quantum world, and it is indeed possible to have quantum error correcting codes.

{\em Conclusions.--} It is possible to encode classical information in shared quantum states in such a way that the information is not in the reduced states of the subsystems, but only in the correlations. The question that we ask in this paper is whether the same can be possible for quantum information -- can quantum information be ``masked'', i.e., encoded only in the correlations? Interestingly, it turns out that while this is in general not possible, i.e., it is not possible to mask arbitrary quantum states, quantum information in certain restricted sets of states, that contain nonorthogonal states, can be masked. The results are in a certain sense complementary to no-cloning and no-deleting, as cloning and deleting are possible only for orthogonal quantum states.

However, if we allow for more than two parties, i.e., $A$, $B$, $C$ and so on, then it is \emph{possible} to mask an arbitrary quantum state. A straightforward example of this is to use an error correction code~\cite{QECC}. However, collusion between any two parties would then reveal the encoded quantum information, at least in part. This has important implications for quantum interacting provers scenarios~\cite{Fitzsimons}. In other words, the goal of quantum error correction is to store all quantum information in correlation. Therefore, the no-go theorem here fundamentally limits the amount and the flavour of quantum information that can be stored bipartite quantum correlations.

Moreover, our masking protocol forms the basis for quantum \emph{secret sharing}~\cite{secretsharing, besh}. Quantum mechanics allows for secret sharing of classical information from a so-called ``boss'' to her ``subordinates'', such that the subordinates are unable to retrieve the information without collaboration between themselves. It is clear that the states chosen by the boss to encode the secret classical bit, and send to her subordinates, can be from a set of orthogonal quantum states that can be masked, as masked information cannot be decoded by the subordinates by local quantum operations without classical communication. Similarly, if the boss wants to send quantum information to her subordinates, she has to choose from a set of quantum states, which in general, will not be orthogonal. The results obtained here can therefore be used to choose the substrates for secret sharing of classical or quantum information. 

The analysis of the sets of states that can be masked reveals that quantum information stored strictly in the phases can always be masked. This is interesting from the perspective that it is the phase of the quantum state that is considered to be the quintessentially quantum aspect, and for example leads to quantum interference, and it is exactly this phase that can be masked just like classical information. Quantum states having information only in the phases falls on a hyper-disk. The fact that such quantum states can be masked reminds us of other quantum information strategies and results like remote state preparation~\cite{eita-remote, chb}, measurement-based quantum computation~\cite{eita-cluster}, the no universal-NOT gate~\cite{eita-NOT}, and parallel and anti-parallel states~\cite{eita-GP, sen2}.

In this respect, it is interesting to uncover whether there can be a (probabilistic) mixture of two orthogonal mixed multipartite states so that there is no information available about the probability when the mixture is accessed locally. However, there will still be a classical bit that will be hidden (``locally-masked''), if this question is answered in the affirmative. It is also interesting to know if there can be a set of superposed states of three orthogonal pure multiparty states so that there is no information available about the (complex) superposition coefficients when an arbitrary element of the set is accessed locally? If true, this will be local-masking of a qutrit.

The no-masking theorem imply that quantum qubit commitment -- of which quantum bit commitment is a dephased version -- is not possible. We have also discussed the potential of using the sets of maskable sets as substrates for secret sharing of classical and quantum information. It is also possible to see that one can consider variations of the maskers considers here, in particular as partial maskers, local maskers, and stochastic approximate maskers. Our results will have important applications in quantum communication and quantum information protocols that require hiding of information in composite quantum systems.

\begin{acknowledgements}
{\bf Acknowledgements.} 
KM thanks S. Bandyopadhyay, B. Hunt, F. Pollock for discussions. We thank J. Fitzsimons for pointing out the connection to quantum error correction codes and masking quantum information. KM thanks the Harish-Chandra Research Institute (HRI), Allahabad for hospitality during the development of these ideas. The ideas in this paper took shape during two meetings at HRI, in 2011 and 2015 (QIPA-2011 and QIPA-2015).
\end{acknowledgements}


\begin{thebibliography}{50}
\bibitem{eita-spooky} This is after the phrase ``spooky action at a distance'' coined by A. Einstein~\cite{eita-o-spooky}.
\end{thebibliography}
\end{document}